\documentstyle[12pt,bezier]{article}
\begin{document}

\title{The Quantum Adiabatic Approximation and~the~Geometric~Phase}
\author{Ali Mostafazadeh\thanks{E-mail: alimos@phys.ualberta.ca}\\ \\
Theoretical Physics Institute, University of Alberta, \\
Edmonton, Alberta,  Canada T6G 2J1.}
\date{June 1996}
\maketitle

\begin{abstract}
A precise definition of an adiabaticity parameter $\nu$ of a 
time-dependent Hamiltonian is proposed. A variation of the 
time-dependent perturbation theory is presented which yields a
series expansion of the evolution operator $U(\tau)=\sum_\ell 
U^{(\ell)}(\tau)$ with $U^{(\ell)}(\tau)$ being at least of the order 
$\nu^\ell$. In particular $U^{(0)}(\tau)$ corresponds to the adiabatic 
approximation  and  yields Berry's  adiabatic phase. It is shown 
that this series expansion has nothing to do with the 
$1/\tau$-expansion of $U(\tau)$. It is also shown that the
non-adiabatic part of the evolution operator is generated
by a transformed Hamiltonian which is off-diagonal in the
eigenbasis of the initial Hamiltonian. Some related issues
concerning the geometric phase are also discussed.
\end{abstract}

\baselineskip=18pt

\newpage

\section{Introduction}
Since the introduction of the adiabatic geometrical phase by Berry 
\cite{berry1984}, the study of the cyclic evolution of quantum states 
of non-conservative quantum systems (explicitly time-dependent
Hamiltonians) has attracted much attention.  By definition a pure 
cyclic state $|{\mbox{\tiny${\cal N}$}};\tau\rangle\langle{\mbox{\tiny${\cal N}$}};\tau|$ is an eigenstate of the 
evolution operator $U(\tau)$, i.e.,
	\begin{equation}
	U(\tau)=:\sum_{\mbox{\tiny${\cal N}$}} e^{i\beta_{\mbox{\tiny${\cal N}$}}(\tau)}
	|{\mbox{\tiny${\cal N}$}};\tau\rangle\langle{\mbox{\tiny${\cal N}$}};\tau|\:, ~~~~~~\beta_{\mbox{\tiny${\cal N}$}}(\tau)\in[0,2\pi)\:.
	\label{exact-u}
	\end{equation}
Therefore the quantity of main importance in the study of cyclic states
and the accompanying phases $\beta_{\mbox{\tiny${\cal N}$}}(\tau)$ is the evolution 
operator $U(\tau)$. This operator is defined by 
	\begin{equation}
	|\psi(\tau)\rangle=:U(\tau)\,|\psi(0)\rangle\;,
	\nonumber
	\end{equation}
where $|\psi(\tau)\rangle$ is the solution of the Schr\"odinger equation. 
Alternatively,  one can define $U(\tau)$ as the solution of
	\begin{eqnarray}
	\frac{d}{d\tau}\, U(\tau)&=&-\frac{i}{\hbar}\:
	H(\tau)\,U(\tau)
	\label{schrodinger}\\
	U(0)&=&1\;,\nonumber
	\end{eqnarray}
where $H=H(\tau)$ stands for the Hamiltonian and $\tau\in[0,\infty)$ 
parameterizes the time. The solution of  Eq.~(\ref{schrodinger})  can
be implicitly expressed in the form:
	\begin{equation}
	U(\tau):={\cal T}\:e^{-\frac{i}{\hbar}\int_0^\tau H(t)dt}\;,
	\label{q1}
	\end{equation}
where ${\cal T}$ denotes the time-ordering operator \cite{bohm-qm}. 

The purpose of this article is two fold. First, a precise definition
of an {\em adiabaticity parameter} $\nu$ will be given. This is a 
parameter which quantifies the rapidity of the time-dependence of 
the Hamiltonian. Next, a series expansion of the evolution operator 
$U(\tau)$ will be proposed whose $\ell$-th term is at least of the
order $\nu^\ell$. In particular the adiabatic approximation:
	\begin{equation}
	U(\tau)\approx\sum_ne^{i\alpha_n(\tau)}\, |n;\tau\rangle
	\langle n;0|\;,
	\label{q2}
	\end{equation}
is recovered as the first term. In Eq.~(\ref{q2}), $|n;t\rangle$ are the
instantaneous eigenstate vectors of the Hamiltonian, i.e.,
	\begin{equation}
	H(t)|n;t\rangle=E_n(t)\,|n;t\rangle\;,
	\label{q4}
	\end{equation}
and 
	\begin{equation}
	\alpha_n(t):=\delta_n(t)+\gamma_n(t)\,,~~~
	\delta_n(t):=-\frac{1}{\hbar}\int_0^t E_n(t')dt'\,,~~~
	\gamma_n(t):=i\int_0^t\langle n;t'|\frac{d}{dt'}|n;t'\rangle\:dt'\;.
	\label{q5}
	\end{equation}
If  the Hamiltonian is $T$-periodical, i.e., $H(T+t)=H(t)$, the $\alpha_n(T),
~\delta_n(T)$, and $\gamma_n(T)$ are called {\em adiabatic total
phase angle}, {\em adiabatic dynamical phase angle}, and
{\em adiabatic geometrical phase} or {\em Berry phase angle},  respectively
\cite{berry1984}.

The approach pursued in this article clarifies the misconception
that the adiabatic approximation of $U(\tau)$ is given by the zero-th
order term in its $1/\tau$-expansion, i.e., it is only valid for $\tau\to\infty$. 
It also provides a convenient framework to perturbatively compute 
the non-adiabatic corrections to Berry's adiabatic phase.

\section{The Adiabaticity Parameter}

Following Berry~\cite{berry1984}, consider a parameter-dependent
Hamiltonian:
	\begin{equation}
 	H[R]=\sum_n E_n[R]\: |n;R\rangle\langle n;R|\;,
	\label{h[r]}
	\end{equation}
where $R=(R^1,\cdots,R^d)$ are real parameters viewed as the 
coordinates of a parameter space ${\cal M}$, $E_n[R]$ are the
real eigenvalues and $\{|n;R\rangle\}$ is a complete orthonormal set of
eigenvectors of $H[R]$. In this case a time-dependent Hamiltonian 
$H(t)$ corresponds to a curve $C:[0,\tau]\to {\cal M}$ in ${\cal M}$:
	\[ H(t):=H[C(t)]=H[R_C(t)]\:,~~~~{\rm with}~~C(t)=:\left( R^1_C(t),
	\cdots,R_C^d(t)\right)=:R_C(t)\;.\]
Furthermore, suppose that the spectrum of $H[R]$ for all $R\in{\cal M}$
is discrete,  the energy eigenvalues $E_n[R]$ are non-degenerate, and 
there is no level-crossing, i.e., for every $t\in [0,\tau]$, $E_n[R_C(t)]<
E_{n+1}[R_C(t)]$. 

In order to define a dimensionless adiabaticity parameter, first one 
defines a {\em characteristic frequency}:
	\begin{equation}
	\omega_c(\tau_1,\tau_2):={\rm Sup}\left\{ |A_{mn}(t)|\;:\;
	n\neq m=0,1,\cdots\;,~~t\in[\tau_1,\tau_2]
	 \right\} \;;~~~~[\tau_1,\tau_2]\subseteq [0,\tau]\;,
	\label{omega_c}
	\end{equation}
with ``Sup'' abbreviating supremum (least upper bound) and
	\begin{eqnarray}
	A_{mn}(t)&:=&\langle m;t|\frac{d}{dt}|n;t\rangle
	\label{a}\\
	&=&\frac{\langle m;t|\left[ \frac{d}{dt}H(t)\right]
	|n;t\rangle}{E_n(t)-E_m(t)}
	\;,~~~~~{\rm for}~~m\neq n\;.
	\label{a-m<>n}
	\end{eqnarray}
Then the desired {\em adiabaticity parameter} is defined 
according to
	\begin{equation}
	\nu:=\frac{\hbar\:\omega_c(0,\tau)}{\Delta E}\;,
	\label{slow}
	\end{equation}
where $\Delta E$ is a convenient energy scale. For example one 
can take $\Delta E$ to be the first  transition energy of the initial 
Hamiltonian, i.e., $\Delta E:=E_1(0)-E_0(0)$.

The main motivation for this definition is the fact that according to
Eqs.~(\ref{omega_c}), (\ref{a-m<>n}), and (\ref{slow}), $\omega_c(0,\tau)$ 
and therefore $\nu$ involve time derivatives of the Hamiltonian $H(t)$. 
The use of $\omega_c(0,\tau)$ in the definition of the adiabaticity parameter 
will be self-evident  once one examines the evolution operator. The role of 
$\Delta E$ is to provide a convenient energy ($\hbar\times$frequency)
scale. 

It  must also be  emphasized that by definition, $\nu$ 
is a ``global'' quantity which characterizes the time-dependence of the 
Hamiltonian. In particular,  it is neither equal nor proportional to $1/\tau$. 
Furthermore, note that although $|n;t\rangle$ are only determined up to  arbitrary 
$R_C(t)$-dependent phase factors (gauge transformations along 
$C$, \cite{bohm-qm}), $|A_{mn}(t)|$ with $m\neq n$ and therefore
$\omega_c(\tau_1,\tau_2)$ and $\nu$ are independent of the choice 
of such phases (they are gauge-invariant quantities). In other words, 
$\nu$ is well-defined.

\section{Computation of $U(\tau)$}
Consider the definition of the time-ordered product in (\ref{q1}):
	\begin{equation}
	U(\tau):=\lim_{N\to\infty}\:
	\left[ 1-\frac{i}{\hbar}H(t_N)\epsilon\right] \cdots
	\left[ 1-\frac{i}{\hbar}H(t_k)\epsilon\right] \cdots
	\left[ 1-\frac{i}{\hbar}H(t_0)\epsilon\right]\;,
	\label{time-ordered}
	\end{equation}
where $\epsilon :=\tau/N$ and $t_k:=k\epsilon$, for $k=0,1,\cdots N$.
Using the orthonormality and completeness of the energy eigenvectors
$|n;t\rangle$, one can compute:
	\begin{eqnarray}
	U(\tau)&=&\lim_{N\to\infty}\:
	\sum_{n_0,\cdots,n_N} \left\{
	|n_N;t_N\rangle\langle n_N;t_N|
	\left[ 1-\frac{i}{\hbar}H(t_N)\epsilon\right] 
	|n_{N-1};t_{N-1}\rangle\langle n_{N-1};t_{N-1}|\right. \nonumber\\
	&&\cdots\nonumber\\
	&&|n_k;t_k\rangle\langle n_k;t_k|
	\left[ 1-\frac{i}{\hbar}H(t_k)\epsilon\right] 
	|n_{k-1};t_{k-1}\rangle\langle n_{k-1};t_{k-1}|\nonumber\\
	&&\cdots\nonumber\\
	&&\left. |n_1;t_1\rangle\langle n_1;t_1|
	\left[ 1-\frac{i}{\hbar}H(t_0)\epsilon\right] 
	|n_0;t_0\rangle\langle n_0;t_0| \right\}\;,\nonumber\\
	&=&\lim_{N\to\infty}\:
	\sum_{n_0,\cdots,n_N}\left(
	e^{-\frac{i}{\hbar}\sum_{j=0}^NE_{n_j}(t_j)\:\epsilon}\: \prod_{k=1}^{N}
	\langle n_k;t_k|n_{k-1};t_{k-1}\rangle\right) |n_N;t_N\rangle\langle n_0;t_0|
	\;.\nonumber
	\end{eqnarray}
Introducing 
	\begin{eqnarray}
	K_{n_Nn_0}(\tau)&:=&\langle n_N;\tau|U(\tau)|n_0;0\rangle\nonumber\\
	&=&
	\lim_{N\to\infty}\:\sum_{n_1,\cdots,n_{N-1}}\: 
	e^{-\frac{i}{\hbar}\sum_{j=1}^{N-1}E_{n_j}(t_j)\:\epsilon}\: \prod_{k=1}^{N}
	\langle n_k;t_k|n_{k-1};t_{k-1}\rangle\;,
	\label{q7}
	\end{eqnarray}
one then has:
	\begin{equation}
	U(\tau)=\sum_{mn}\: K_{mn}(\tau)\:|m;\tau\rangle\langle n;0|\;.
	\label{q6}
	\end{equation}
The computation of the terms in the product in Eq.~(\ref{q7}), is straightforward:
	\begin{eqnarray}
	\langle n_k;t_k|n_{k-1};t_{k-1}\rangle&=&\delta_{n_kn_{k-1}}-\epsilon\:
	\left. \langle n_k;t|\frac{d}{dt}|n_{k-1};t\rangle\right|_{t=t_{k-1}}
	+{\cal O}(\epsilon^2)\;,\nonumber\\	
	&=& e^{-\epsilon A_{n_kn_{k-1}}(t_{k-1})}\delta_{n_kn_{k-1}}+
	\epsilon \:\Delta_{n_kn_{k-1}}(t_{k-1}) +{\cal O}(\epsilon^2)\;,\nonumber\\
	&=& e^{-\epsilon A_{n_kn_{k}}(t_{k})} \left[ \delta_{n_kn_{k-1}}+
	\epsilon \:e^{\epsilon A_{n_kn_{k}}(t_{k})}\Delta_{n_kn_{k-1}}(t_{k})
	\right] + {\cal O}(\epsilon^2)\;,
	\label{q8}
	\end{eqnarray}
where
	\begin{equation}
	\Delta_{mn}(t):=  (\delta_{mn}-1)A_{mn}(t)\;.
	\label{delta}
	\end{equation}

Substituting (\ref{q8}) in (\ref{q7}), one obtains $2^N$ terms which can be 
arranged in the order of the appearance of different powers of $\epsilon$. 
In this way one finds only $N$ terms of order $\epsilon^0=1$.\footnote{Note 
that a sum of $N^\ell$ terms of order  $\epsilon^\ell$ is of order $\epsilon^0
=1$.} These will be denoted by $K_{mn}^{(\ell)}(\tau)$:
	\begin{equation}
	K_{mn}(\tau)=\lim_{N\to\infty}\sum_{\ell=0}^N 
	K^{(\ell)}_{mn}(\tau)\;.
	\label{cun}
	\end{equation}
Performing the algebra one finds:
	\begin{eqnarray}
	K^{(\ell)}_{n_Nn_0}(\tau)&:=&\sum_{n_1\cdots n_{N-1} }~~
	\sum_{t_{i_1}<\cdots<t_{i_{\ell}}=0}^\tau \epsilon^\ell 
	\left[ e^{\frac{i}{\hbar}\sum_{j=1}^N
	\left[ -E_{n_j}(t_j)+i\hbar A_{n_jn_j}(t_{j}) \right]
	\epsilon } \;\;\; \times\right. \nonumber\\
	&& 	\delta_{n_0n_1}\cdots \delta_{n_{i_1-2}n_{i_1-1}}
	e^{\epsilon A_{n_{i_1}n_{i_1}}(t_{i_1})}
	\Delta_{n_{i_1}n_{i_1-1}}(t_{i_1})
	\delta_{n_{i_1}n_{i_{1}+1}} \cdots  \nonumber\\
	&& \left.\delta_{n_{i_\ell-2}n_{i_\ell-1}}
	e^{\epsilon A_{n_{i_{\ell}}n_{i_{\ell}}}(t_{i_{\ell}})}
	\Delta_{n_{i_{\ell}}n_{i_{\ell}-1}}(t_{i_{\ell}})
	\delta_{n_{i_{\ell}}n_{I_{\ell}+1}}\cdots \delta_{n_{N-1}n_N}
	\right] +{\cal O}(\epsilon)\;,
	\label{q10}
	\end{eqnarray}
where $\ell=0,1,\cdots,N$ and $t_k=k\epsilon/N=0,\epsilon,\cdots,\tau$. 
In particular,
	\begin{eqnarray}
	K^{(0)}_{n_Nn_0}(\tau)&=&\Gamma_{n_N}(\tau)\:
	\delta_{n_Nn_0}+{\cal O}(\epsilon)\;,
	\label{k0}\\
	K^{(1)}_{n_Nn_0}(\tau)&=&-\frac{i}{\hbar}\:
	\Gamma_{n_N}(\tau)\sum_{t_i=t_j} ^{t_k} \epsilon\: H'_{n_Nn_0}(t_i)
	+{\cal O}(\epsilon)\;,\nonumber\\
	&=&-\frac{i}{\hbar}\sum_m\left[ K^{(0)}_{n_Nm}(\tau)
	\sum_{t_i=0} ^{\tau} \epsilon\: H'_{mn_0}(t_i)\right]
	+{\cal O}(\epsilon)\;,
	\label{k1}
	\end{eqnarray}
where
	\begin{eqnarray}
	\Gamma_n(t_k)&:=&e^{\frac{i}{\hbar}\sum_{t_j=0}^{t_k}
	\left[  -E_{n}(t_j)+i\hbar A_{nn}(t_{j}) \right]
	\epsilon } \;, 
	\label{Gamma}\\
	H'_{mn}(t_k)&:=&i\hbar\:\Gamma_{m}^*(t_{k}) 
	\Delta_{mn}(t_{k})\Gamma_{n}(t_{k})\;,
	\label{h^1_mn}
	\end{eqnarray}
and $\Gamma_m^*(t_k)=1/\Gamma_m(t_k)$ is the complex conjugate
of $\Gamma_m(t_k)$.

Furthermore, for every $\ell>1$, one can express $K^{(\ell)}_{mn}(\tau)$ 
in terms of $K^{(0)}_{mn}(\tau)$ and $H'_{mn}(t_i)$, namely	\begin{eqnarray}
	K^{(\ell)}_{m~m_\ell}(\tau)&=&\left( \frac{-i}{\hbar}\right)^\ell
	\sum_{t_{i_1}<\cdots<t_{i_{\ell}}=0}^\tau\epsilon^\ell
	\sum_{m_0\cdots m_{\ell-1}} K^{(0)}_{m~m_0}(\tau)
	\left[ \prod_{a=1}^{\ell} H'_{m_{a-1}m_a}(t_{i_a}) \right]+
	{\cal O}(\epsilon)\;,\nonumber\\
	&=& 
	\frac{1}{\ell\, !} \left( \frac{-i}{\hbar}\right)^\ell
	\sum_{m_0\cdots m_{\ell-1}}K^{(0)}_{m~m_0}(\tau)
	\sum_{t_{i_1}\cdots t_{i_{\ell}}=0}^\tau\epsilon^\ell\:
	{\cal T}\left[ \prod_{a=1}^{\ell} H'_{m_{a-1}m_a}(t_{i_a}) \right]+
	{\cal O}(\epsilon)\;.
	\label{k^l}
	\end{eqnarray}
Therefore if one defines:
	\begin{eqnarray}
	U^{(\ell)}(\tau)&:=& \lim_{N\to\infty}\sum_{mn}K^{(\ell)}_{mn}(\tau)
	|m;\tau\rangle\langle n;0|\;,
	\label{u^l}\\
	H'(t)&:=&\lim_{N\to\infty}\sum_{mn}H'_{mn}(t)|m;0\rangle
	\langle n;0|\;,\nonumber\\
	&=&i\hbar\sum_{mn} e^{-i[\alpha_m(t)-\alpha_n(t)]}\, \Delta_{mn}(t)
	\,|m;0\rangle\langle n;0| \;,\nonumber\\
	&=&-i\hbar\sum_{m\neq n} e^{-i[\alpha_m(t)-\alpha_n(t)]}\, A_{mn}(t)
	\,|m;0\rangle\langle n;0| \;,
	\label{h^1}
	\end{eqnarray}
then
	\begin{equation}
	U(\tau)=\sum_{\ell=0}^\infty U^{(\ell)}(\tau)=
	U^{(0)}(\tau)\: \left[ {\cal T}e^{-\frac{i}{\hbar}\int_0^\tau dt\, 
	H'(t)}\right]\;,
	\label{u=T...}
	\end{equation}
where by taking $N\to \infty$ the sums of the form $\sum_{t_i=t}^{t'}
\epsilon~f(t_i)$ have been promoted to the integrals $\int_t^{t'} dt\,f(t)$.
In particular one has:
	\begin{eqnarray}
	U^{(0)}(\tau)&=&\sum_ne^{i\alpha_n(\tau)}\, |n;\tau\rangle
	\langle n;0|\;,
	\label{u0}\\
	U^{(1)}(\tau)&=&U^{(0)}(\tau)\left[\frac{-i}{\hbar}\int_0^\tau 
	dt\,H'(t)\right]\;.
	\label{u^1=}
	\end{eqnarray}

By construction  $U^{(0)}(\tau)$ yields $U(\tau)$ in the adiabatic 
approximation (\ref{q2}).Therefore, adiabatic approximation is valid 
only if one can neglect  $\int_o^\tau dt\, H'(t)$ in (\ref{u=T...}). It is 
not difficult to observe that this condition is fulfilled if the adiabaticity 
parameter $\nu$ as defined by Eq.~(\ref{slow}) is negligible. 
In fact, for every $t_1$ and $t_2$ satisfying $0\leq t_1\leq t_2\leq\tau$, 
one has:
	\begin{equation}
	|\langle m;0|\int_{t_1}^{t_2}dt\: H'(t)|n;0\rangle|\leq
	\int_{t_1}^{t_2} dt\: |\langle m;0|H'(t)|n;0\rangle|\leq\hbar
	\int_{t_1}^{t_2} dt\: |A_{mn}(t)|\leq \, \Delta\tau\:\Delta E
	\:\nu\;,
	\label{h^1<}
	\end{equation}
where $\Delta\tau:=\tau_2-\tau_1$ with $[\tau_1,\tau_2]\subseteq
[0,\tau]$ is the time interval over which $\omega_c$ is non-vanishing,
i.e., $\tau_1$ and $\tau_2$ and therefore $\Delta\tau$ are defined by 
the condition: 
	\begin{equation}
	\omega_c(\tau'_1,\tau'_2)=0~~~~{\rm if~and~only~if }~~~~ \tau'_2<
	\tau_1~~{\rm or}~~ \tau'_1>\tau_2\;. 
	\label{condition}
	\end{equation}
Note that usually $\Delta\tau$ is a finite  time interval whereas $\tau$ 
may be infinite (arbitrarily large).  A case  for which $\Delta\tau=\tau$ 
is when the Hamiltonian depends periodically on time. In this case 
however, the physically interesting features of the evolution is given 
by $\Delta\tau=\tau\leq T$, where $T$ is the period of  the Hamiltonian.
Particularly interesting is the case $\tau=T$. 

In general, $\Delta\tau$ may become arbitrarily large in which case
(\ref{h^1<}) is a trivial statement. However, even in this case $\nu$ 
plays a most  important role. In fact, one can argue that in general
 the following relation holds:
	\begin{equation}
	U(\tau)=\sum_{\ell=0}^{N-1} U^{(\ell)}(\tau)+{\cal O}(\nu^N)\;.
	\label{N-approximation}
	\end{equation}
In particular the adiabatic approximation is valid if and only if $\nu\ll 1$.
In order to establish Eq.~(\ref{N-approximation}), it is 
sufficient\footnote{This is because $U^{(\ell)}(\tau)$ involves
$\ell$ copies of $H'(\tau)$ in its integrand.} to show
that $U^{(1)}(\tau)$ and therefore $\int dt\: H'(t)$ is at least of order 
$\nu$. This is however self-evident since one knows that if $\nu=0$,
then $U(\tau)=U^{(0)}(\tau)$. Consequently, in a $\nu$-expansion
of $U(\tau)$, $U^{(1)}(\tau)$ is necessarily of order $\nu$ or higher.

An alternative way of expressing Eq.~(\ref{N-approximation}) is to 
define  $\tilde U^{(\ell)}(\tau):=U^{(\ell)}(\tau)/\nu$ and $\tilde H'
(\tau):=H'(\tau)/\nu$ and write $U(\tau)$ in the form:
	\begin{equation}
	U(\tau)=\sum_{\ell=0}^\infty \tilde U^{(\ell)}(\tau)\:
	\nu^\ell=U^{(0)}(\tau)\:\left[ {\cal T}
	e^{-\frac{i\nu}{\hbar}\int_0^\tau dt\:
	\tilde H'(t)}\right] \;.
	\label{nu-expansion}
	\end{equation}
This equation must not however be viewed as a true power series
expansion of $U(\tau)$ in $\nu$, for $\tilde U^{(\ell)}(\tau)$ may 
in general depend  on $\nu$. 

A simple consequence of Eq.~(\ref{nu-expansion}) is the fact that
the essential ingredient which determines $U(\tau)$ beyond the 
adiabatic approximation is not $\tau$, but $(\tau_1,\tau_2)$ of  
(\ref{condition}) and $\nu$. For physically realistic non-periodic 
Hamiltonians, the latter quantities take finite values whereas the 
duration $\tau$ of the evolution of the system may be arbitrarily large. 
For a quantum system with a  periodic Hamiltonian $H(t)=H(t+T)$, the 
physically interesting case is when $\tau_2=\tau\leq T$. In this case, 
$\nu$ may or may not be determined by $T$. This is because, in 
general the parameters $R^\mu_C(t)$ of the Hamiltonian have 
different  periods. These are necessarily of the form $T_\mu=T/z_\mu$, 
respectively,  where $z_\mu$ is a positive integer. Clearly 
$\omega_c(0,\tau)$ and consequently $\nu$ will depend on the largest 
value of $z_\mu$. Therefore, in general  the statement that ``the adiabatic 
approximation is valid if the period of the Hamiltonian is large,'' is false.
The only case where $\nu$ depends on $T$ and the preceding
statement is valid, is the case where the energy eigenstates are 
time-dependent and either there is effectively one changing parameter or 
$T_\mu=T$ for all $\mu$. Typical examples of these
two cases are a  spin in a precessing magnetic field \cite{bohm-qm} 
and a spin in a precessing and nutating magnetic field with equal 
precession and nutation periods, respectively. 

Another useful observation is that in the eigenbasis $\{|n;0\rangle\}$ of $H(0)$,
$H'(t)$ has no diagonal matrix elements, i.e.,
	\begin{equation}
	\langle m;0|H'(t)|n;0\rangle =\left\{ \begin{array}{ccc}
	0& {\rm for} & m=n\\ i\hbar\:\mbox{\large$
	\frac{\langle m;t| [\frac{d}{dt}H(t)]|n;t\rangle}{E_m(t)-E_n(t)}\:
	e^{-i[\alpha_m(t)-\alpha_n(t)]}$}&{\rm for}& m\neq n\;,
	\end{array}\right.
	\label{delta1}
	\end{equation}
where use is made of Eqs.~(\ref{h^1}) and (\ref{a-m<>n}). 
This equation also implies that the matrix elements  $\langle m;\tau|
U^{(\ell)}(\tau)|n;0\rangle$ are directly related to $\left( \frac{1}{|n-m|}
\right)^\ell$, i.e., the main contribution to $U^{(\ell)}(\tau)$ comes from 
the nearest energy levels. 

\section{Relation between $H'(t)$ and $H(t)$}

Consider Eq.~(\ref{h^1}). Since the $m=n$ term is missing in the
sum in (\ref{h^1}), one can express $H'(t)$ in the form:	\begin{equation}
	H'(t)=-i\hbar\,\sum_{m\neq n}\left[ \langle m;t|e^{-i\alpha_m(t)}\right]
	\frac{d}{dt}\left[ e^{i\alpha_n(t)}|n;t\rangle\right]\:|m;0\rangle\langle n;0| \;.
	\end{equation}
This equation together with the definition of $U^{(0)}(t)$ leads to
	\begin{equation}
	H'(t)=U^{(0)\dagger}(t)\:H(t)\:U^{(0)}(t)-i\hbar\,U^{(0)\dagger}(t)
	\frac{d}{dt}\,U^{(0)}(t)\;.
	\label{gauge}
	\end{equation}
The resemblance of this equation to the gauge transformations of the 
gauge potential in non-Abelian gauge theories (connections on 
principal fiber bundles) is remarkable, \cite{nakahara}.  Moreover, a 
simple calculation shows that
	\begin{eqnarray}
	U^{(0)\dagger}(t)\:H(t)\:U^{(0)}(t)&=&\sum_{n}E_n(t)\:
	|n;0\rangle\langle n;0|\;,
	\label{1st}\\
	H'(t)&=&-i\hbar\,\sum_{n\neq m} \langle m;0|U^{(0)\dagger}(t)
	\frac{d}{dt}\,U^{(0)}(t)|n;0\rangle\: |m;0\rangle\langle n;0|\;.
	\label{2nd}
	\end{eqnarray}
Therefore, the role of the first term on the right hand side of
(\ref{gauge}) is to cancel the diagonal matrix elements of the second
term.

Eqs.~(\ref{gauge}) and (\ref{2nd}) signify the importance of the 
adiabatic approximation in the determination of the non-adiabatic 
corrections $U^{(\ell)}(\tau)$ ($\ell>0$) to $U^{(0)}(\tau)$ and consequently
the exact evolution operator $U(\tau)$. 

In order to fully appreciate the meaning of Eq.~(\ref{gauge}), one 
should recall the effect of a general time-dependent unitary transformation
of  the Hilbert  space ${\cal H}$. Let ${\cal U}(t):{\cal H}\to{\cal H}$ 
be such a  transformation and suppose that the transformed state vectors
	\begin{equation}
	|\check\psi(t)\rangle:={\cal U}(t)\:|\psi(t)\rangle\;,
	\label{psi-trans}
	\end{equation}
satisfy the following Schr\"odinger equation:
	\begin{equation}
	i\hbar\:\frac{d}{dt}\:|\check\psi(t)\rangle=\check H(t)\,|\check\psi(t)\rangle\;.
	\label{schro-trans}
	\end{equation}
Then requiring $|\psi(t)\rangle$ to satisfy the Schr\"odinger equation
defined by the original Hamiltonian $H(t)$, one finds:
	\begin{equation}
	\check H(t)={\cal U}(t)\, H(t)\,{\cal U}^\dagger(t)-
	i\hbar\:{\cal U}(t)\,\frac{d}{dt}\,{\cal U}^\dagger(t)\;.
	\label{h-trans}
	\end{equation}
Therefore Eq.~(\ref{gauge}) is a particular example of Eq.~(\ref{h-trans})
with ${\cal U}(t)=U^{(0)\dagger}(t)$ and $\check H(t)=H'(t)$. In other words,
the non-adiabatic effects are given by a transformed Hamiltonian $H'(t)$,
where the transformation is performed by the inverse of the adiabatically
approximate evolution operator, i.e.,  $U^{(0)\dagger}(t)$.

Transformation (\ref{psi-trans}) may be used to yield an exact solution
of the Schr\"odinger equation, if one can find a ${\cal U}(t)$ that makes
the transformed Schr\"odinger equation (\ref{schro-trans}) easily solvable.
The Rabi-Ramsey-Schwinger method \cite{rabi} is a special case of
this approach where one uses the symmetry of the problem to find
a unitary transformation ${\cal U}(t)$ which renders $\check H$ 
time-independent.\footnote{For a recent review of the application of this
method to a spin system in a precessing magnetic field, see 
Ref.~\cite{bohm-qm}.} This is in general possible for  cranked Hamiltonians
with a fixed cranking direction \cite{wang,p7}. 

\section{Time-Dependent Unitary Transformations and the Geometric Phase}

There is another significant application of time-dependent unitary
transformations of the Hilbert space when the Hamiltonian is $T$-periodic.
In this case, one knows from the Floquet theory \cite{moore} that
the evolution operator is given by $U(\tau)=Z(\tau)\exp[-i\tau\tilde H/\hbar]$,
where $Z(\tau)$ is a $T$-periodic unitary operator with $Z(T)=Z(0)=1$,
and $\tilde H$ is a time-independent Hermitian operator. Then by definition
the cyclic states $|{\mbox{\tiny${\cal N}$}};T\rangle\langle{\mbox{\tiny${\cal N}$}};T|$ at $\tau=T$ are given by the 
eigenstates of $\tilde H$ and the associated total phase angles are 
$\beta_{\mbox{\tiny${\cal N}$}}(T)=-T\langle{\mbox{\tiny${\cal N}$}};T|\tilde H|{\mbox{\tiny${\cal N}$}};T\rangle/\hbar$. Next consider choosing
${\cal U}(t)$ of Eq.~(\ref{psi-trans}) to be $Z^\dagger(t)$ and $|\psi(0)\rangle=
|\check\psi(0)\rangle=|{\mbox{\tiny${\cal N}$}};T\rangle$. Then by virtue of Eq.~(\ref{h-trans}), one has:
	\begin{eqnarray}
	\beta_{\mbox{\tiny${\cal N}$}}(T)&=&-\frac{T}{\hbar} \langle\psi(0)|\tilde H|\psi(0)\rangle=
	T\left[ -\frac{1}{\hbar}\langle\psi(t)|H(t)|\psi(t)\rangle+i\langle\psi(0)|Z^\dagger(t)\:
	\frac{d}{dt}\:Z(t)|\psi(0)\rangle\right]\:,\nonumber\\
	&=&	\Delta_{\mbox{\tiny${\cal N}$}}(T)+\Gamma_{\mbox{\tiny${\cal N}$}}(T)\;,
	\label{total}\\
	\Delta_{\mbox{\tiny${\cal N}$}}(\tau)&:=&-\frac{1}{\hbar}\int_0^\tau dt\:\langle\psi(t)|H(t)|\psi(t)\rangle\;,
	\label{aa-dyn}\\
	\Gamma_{\mbox{\tiny${\cal N}$}}(\tau)&:=&i\int_0^\tau dt\:\langle\phi_{\mbox{\tiny${\cal N}$}}(t)|\frac{d}{dt}\:
	|\phi_{\mbox{\tiny${\cal N}$}}(t)\rangle\;,
	\label{aa-geo}
	\end{eqnarray}
where $|\phi(t)\rangle:=Z(t)|\psi(0)\rangle$. The phase angles $\Delta_{\mbox{\tiny${\cal N}$}}(\tau)$ and
$\Gamma_{\mbox{\tiny${\cal N}$}}(\tau)$ were first introduced by Aharonov and Anandan \cite{aa},
and called the {\em dynamical} and {\em geometrical} phase angles.\footnote{Note
that in general the Hamiltonian needs not be periodic and the Floquet
theory does not apply. However, even in this case one can define
$|\phi_{\mbox{\tiny${\cal N}$}}(t)\rangle$ and Eqs.~(\ref{total})-(\ref{aa-geo}) are valid.
See \cite{aa,bohm-qm} for more detail.} They generalize their adiabatic
counterparts previously discovered by Berry \cite{berry1984}. Clearly, 
for $\nu\ll 1$, one has: $|{\mbox{\tiny${\cal N}$}};T\rangle\approx|n;0\rangle$, $\Delta_{\mbox{\tiny${\cal N}$}}(T)\approx
\delta_n(T)$, and $\Gamma_{\mbox{\tiny${\cal N}$}}(T)\approx\gamma_n(T)$.

Next consider the general case. If  ${\cal U}(t)$ is also assumed to leave the
original energy eigenstates invariant, i.e., $\check H(t)$ has the same
eigenstates as $H(t)$, then one can show that ${\cal U}(t)$ and $\check H(t)$
must be of the form:
	\begin{eqnarray}
	{\cal U}(t)&=&e^{if(t)}\:{\cal U}_0\;,
	\label{trans-0}\\
	H(t)&\rightarrow&\check H(t)\:=\:H(t)-\hbar\:\frac{df(t)}{dt}\,1\;,
	\label{h-trans-0}
	\end{eqnarray}
where $f=f(t)$ is a real-valued function, ${\cal U}_0$ is a constant unitary 
transformation, and $1$ is the identity operator on ${\cal H}$. Note that 
transformation (\ref{psi-trans}) with ${\cal U}(t)$ given by (\ref{trans-0}) 
leaves all the observables invariant. It shifts the energy eigenvalues:
	\begin{equation}
	E_n(t)\rightarrow E'_n(t)=E_n(t)-\hbar\:\frac{df(t)}{dt}\;,
	\label{e-trans}
	\end{equation}
but does not affect the observable transition energies $E_n(t)-E_m(t)$.
Therefore, one can identify a physical system with the equivalence
class of all the quantum systems which are related with transformations
of the form (\ref{psi-trans},\ref{trans-0}). 

A simple consequence of this observation is that unlike the dynamical
phase angle (\ref{aa-dyn}), the geometric phase angle (\ref{aa-geo})
is a physical quantity \cite{bohm-qm}. This statement however requires
a clarification. One can easily see that (\ref{h-trans-0}) changes the 
dynamical phase angle $\Delta_{\mbox{\tiny${\cal N}$}}(\tau)$. It does nevertheless leave 
the difference of two dynamical phase angles and consequently two total 
phase angles invariant. Therefore,  the dynamical phase angle also carries
physically significant information about the evolving system. In fact,
a close look at the experimental results on the detection of  the geometric
phase \cite{exp} clearly shows that the measurable quantities are
related to the differences of total phase angles. This is usually
overshadowed by the fact that in the best studied system, namely 
the two-level spin ($j=1/2$) system in a precessing magnetic field
\cite{berry1984}, the conventional choice of the Hamiltonian (\ref{3.1})
leads to two cyclic states whose total phase angles differ by a minus sign. 
Therefore their difference is twice one of them and it appears that the 
experiments detect a total phase. In reality however, the experiments always
detect the differences between total phase angles, a quantity which
is invariant under (\ref{h-trans-0}). This can be easily seen if one
uses a complete orthonormal set of cyclic states $\{|{\mbox{\tiny${\cal N}$}};\tau\rangle\}$
to compute the expectation values:
	\begin{eqnarray}
	\langle X(\tau)\rangle&:=&\langle\psi(\tau)|X(0)|\psi(\tau)\rangle\:=\:
	\langle\psi(0)|U^\dagger(\tau)\:X(0)\:U(\tau)|\psi(0)\rangle\;,\nonumber\\
	&=&\sum_{{\mbox{\tiny${\cal N}$}},{\mbox{\tiny${\cal N}$}}'}\langle\psi(0)|{\mbox{\tiny${\cal N}$}};\tau\rangle\langle{\mbox{\tiny${\cal N}$}};\tau|
	U^\dagger(\tau)\:X(0)\:U(\tau)|{\mbox{\tiny${\cal N}$}}',\tau\rangle\langle {\mbox{\tiny${\cal N}$}}';\tau|\psi(0)\rangle\;,
	\nonumber\\
	&=&\sum_{{\mbox{\tiny${\cal N}$}},{\mbox{\tiny${\cal N}$}}'}e^{-i[\beta_{\mbox{\tiny${\cal N}$}}(\tau)-\beta_{{\mbox{\tiny${\cal N}$}}'}(\tau)]}
	\langle\psi(0)|{\mbox{\tiny${\cal N}$}};\tau\rangle\:\langle{\mbox{\tiny${\cal N}$}};\tau|X(0)|{\mbox{\tiny${\cal N}$}}',\tau\rangle\:
	\langle {\mbox{\tiny${\cal N}$}}';\tau|\psi(0)\rangle\;,\nonumber
	\end{eqnarray}
where $X(0)$ is an observable and in the last equality used is made of 
Eq.~(\ref{exact-u}).

Incidentally, it is not too difficult to see that by definition, the 
adiabaticity parameter $\nu$ (\ref{slow}), the operator $H'(t)$ of 
Eq.~(\ref{h^1}), and consequently all the non-adiabatic corrections
$U^{(\ell)}(\tau)$ (with $\ell\geq 1$) to $U^{(0)}(\tau)$ are invariant
under the transformation (\ref{trans-0}). Therefore all of these
quantities signify physically measurable effects.

\section{Application to a Magnetic Dipole in a Magnetic Field}
In this section, the utility of the expansion (\ref{u=T...}) is demonstrated
for the quantum system consisting of a magnetic dipole moment subject to
a magnetic field with changing direction. 

The Hamiltonian\footnote{As discussed in the previous section, the
choice of a Hamiltonian is not unique, i.e., one can add a multiple of
the identity operator to Eq.~({3.1}) without having any physical 
consequences.} of this system is given by:
	\begin{equation}
	H(\theta,\varphi)=b \vec R(\theta,\varphi)\cdot \vec J
	=b(\sin\theta\cos\varphi J^1+\sin\theta\sin\varphi J^2+
	\cos\theta J^3)\:,
	\label{3.1}
	\end{equation}
where $b$ is the Larmor frequency,  $\theta$ and $\varphi$ are
the azimuthal and polar angles in spherical coordinates, respectively,
and $\vec J$ is the angular momentum operator with 
components $J^\mu$, $\mu=1,2,3$. Clearly the parameter manifold is 
the two-dimensional sphere $S^2$ and the time-dependence of the 
Hamiltonian is described by a curve $C:[0,\tau]\to S^2$. Without loss 
of generality one can choose a coordinate system in which $C$ does 
not pass through the south pole. This allows one to work with a single 
coordinate patch of $S^2$ which excludes the south pole.

In this patch one has \cite{bohm-qm}:
	\begin{eqnarray}
	E_n(\theta,\varphi)&=&E_n(0,0)=b\hbar \,n\:,~~~{\rm with}~~
	n=0,\:\pm\frac{1}{2}\,,\:\pm1\,,\:\pm\frac{3}{2}\,,\cdots\;,
	\label{3.3}\\
	|n;(\theta,\varphi)\rangle&=&e^{-\frac{i\varphi}{\hbar}J^3}
	e^{-\frac{i\theta}{\hbar}J^2}e^{\frac{i\varphi}{\hbar}J^3}
	|n;(0,0)\rangle\;, ~~~~~\theta\in[0,\pi),~\varphi\in[0,2\pi)\;.
	\label{3.2}
	\end{eqnarray}
By definition $|n;(0,0)\rangle$ are the eigenvectors of  $H(\theta=0,\varphi=0)=
bJ^3$, i.e., 
	\[ J^3|n;(0,0)\rangle=\hbar n|n;(0,0)\rangle \;.\]

In order to compute the operators $U^{(0)}(\tau)$ and $H'(\tau)$, one 
first calculates
	\begin{eqnarray}
	A_{mn}(t)&=&A_\theta^{(mn)}\:\dot \theta(t)
	+A_\varphi^{(mn)}\:\dot\varphi(t)
	\label{a-t}\\
	A_\theta^{(mn)}
	&:=&\langle m;(\theta,\varphi)|\frac{\partial}{\partial\theta}
	|n;(\theta,\varphi)\rangle\;,\nonumber\\
	&=&\:\frac{i}{\hbar}\left[ \sin\varphi \:\langle J_{mn}^1\rangle_0-
	\cos\varphi\:\langle J_{mn}^2\rangle_0\right] \;,
	\label{a_theta}\\
	A_\varphi^{(mn)}&:=&\langle m;(\theta,\varphi)|
	\frac{\partial}{\partial\varphi}
	|n;(\theta,\varphi)\rangle\;,\nonumber\\
	 &=&i\left[
	m(1-\cos\theta)\delta_{mn}+\frac{1}{\hbar}\sin\theta(\cos\varphi\:
	\langle J_{mn}^1\rangle_0+\sin\varphi\:\langle J_{mn}^2\rangle_0)\right]\;,
	\label{a_phi}
	\end{eqnarray}
where $\langle J_{mn}^\mu\rangle_0:=\langle m;(0,0)|J^\mu|n;(0,0)\rangle$,
$(\theta(t),\varphi(t))=C(t)$ and dot stands for $d/dt$.  In the 
derivation of Eqs.~(\ref{a_theta}) and (\ref{a_phi}) use is made of the 
following identities:
	\begin{eqnarray}
	e^{-\frac{i\varphi}{\hbar}J^3}\:J^2\:e^{\frac{i\varphi}{\hbar}J^3}
	&=&-\sin\varphi\:J^1+\cos\varphi \:J^2\;,\nonumber\\
	e^{\frac{i\theta}{\hbar}J^2}\:J^3\:e^{-\frac{i\theta}{\hbar}J^2}
	&=&\cos\theta\:J^3-\sin\theta\:J^1\;,\nonumber\\
	e^{-\frac{i\varphi}{\hbar}J^3}\:J^1\:e^{\frac{i\varphi}{\hbar}J^3}
	&=&\cos\varphi \: J^1+\sin\varphi\:J^2\;.\nonumber
	\end{eqnarray}

Next  step is the calculation of the phase angles  $\alpha_n(t)$ of
Eq.~(\ref{q5}). These are given by:
	\begin{equation}
	\alpha_n(t)=\delta_n(t)+\gamma_n(t)\,,~~~
	\delta_n(t)=-btn\,,~~~\gamma_n(t)=-n\gamma(t)\;,
	\label{alpha-gamma-0}
	\end{equation}
where
	\begin{equation}
	\gamma(t):=\int_0^tdt'\: (1-\cos\theta)\:\dot\varphi=
	\int_0^{\varphi(t)}(1-\cos\theta)d\varphi\;.
	\label{alpha-gamma}
	\end{equation}
In (\ref{alpha-gamma}), $\varphi(0)$ is set to zero and in the second
integral $\varphi$ is used to parameterize the curve $C$, i.e., $\theta=
\theta(\varphi)$.

Eqs.~(\ref{a-t}-\ref{a_phi}), (\ref{alpha-gamma-0}), and (\ref{alpha-gamma})
together with Eqs.~(\ref{u0}) and (\ref{h^1}) yield:
	\begin{eqnarray}
	U^{(0)}(\tau)&=&\sum_{n}e^{-i[b\tau+\gamma(\tau)]n}
	|n;\tau\rangle\langle n;0|\;,
	\label{u0-spin}\\
	H'(t)&=&\sum_{m\neq n}\left[ H^{'mn}_1(t)\,
	\langle J^1_{mn}\rangle_0+H^{'mn}_2(t)\,\langle J^2_{mn}\rangle_0\right]
	\,| m;0\rangle\langle n;0|\:,
	\label{h^1-spin}
	\end{eqnarray}
where
	\begin{eqnarray}
	H^{'mn}_1(t)&:=&
	 e^{i[bt+\gamma](m-n)}
	(\sin\varphi\: \dot\theta+\sin\theta\cos\varphi\:\dot\varphi)\;,
	\label{i1}\\
	H^{'mn}_2(t)&:=&
	e^{i[bt+\gamma](m-n)}
	(-\cos\varphi\:\dot\theta+\sin\theta\sin\varphi\:\dot\varphi)\;.
	\label{i2}
	\end{eqnarray}

In order to further ease the computation of $H'(\tau)$, one can make
the additional assumption that the energy eigenstates have definite
total angular momentum, i.e., 
	\begin{equation}
	|\vec J|^2|n,(\theta,\varphi)\rangle=j(j+1)\:|n,(\theta,\varphi)\rangle\;,
	~~~n=-j,-j+1,\cdots,j\;.
	\label{J2}
	\end{equation}
This assumption is too restrictive for the applications in 
molecular physics, where one encounters systems with cylinderical
symmetry rather than spherical symmetry \cite{bohm-qm}. For a
magnetic dipole in a classical environment, however, one can safely 
make this assumption. In this case, one can use the well-known
relations \cite{sakurai}: 
	\begin{equation}
	J^{\pm}:=J^1\pm iJ^2\;,~~~J^\pm|n;(0,0)\rangle=\hbar C_{\pm n}
	|n\pm 1;(0,0)\rangle\;,~~~C_m:=\sqrt{(j-m)(j-m+1)}\;,
	\label{j+-}
	\end{equation}
to compute $\langle J_{mn}^1\rangle_0$ and $\langle J_{mn}^2\rangle_0$.
This leads to
	\begin{equation}
	\langle J_{mn}^1\rangle_0=\frac{\hbar}{2}\:(C_n\:\delta_{m\,n+1}+
	C_{-n}\:\delta_{m\,n-1})\;,~~~~\langle J_{mn}^2\rangle_0=
	\frac{-i\hbar}{2}\:(C_n\:\delta_{m\,n+1}-C_{-n}\:\delta_{m\,n-1})\;.
	\label{j12}
	\end{equation}
Substituting Eqs.~(\ref{j12}) and carrying out the necessary algebra
one has:
	\begin{eqnarray}
	A_\theta^{(mn)}&=&\frac{1}{2}\:(e^{i\varphi}C_m\delta_{m\,n-1}
	-e^{-i\varphi}C_n\delta_{m-1\,n})\;,
	\label{a_theta-1}\\
	A_\varphi^{(mn)}&=&i\left[ m(1-\cos\theta)\delta_{mn}+
	\frac{1}{2}\: \sin\theta\: (e^{i\varphi}C_m\delta_{m\,n-1}+
	e^{-i\varphi}C_n\delta_{m-1\,n})\right]\;,
	\label{a_phi-1}\\
	H'(t)&=&\frac{\hbar}{2}\left[ {h'}(t)\sum_nC_n\:
	|n+1;0\rangle\langle n;0|+ {h'}^*(t)\sum_nC_n\:
	|n-1;0\rangle\langle n;0|\right] \;,\nonumber\\
	&=&\frac{\hbar}{2}\left[ {h'}(t)\:e^{-i\frac{\theta_0}{\hbar}J^2}
	\sum_nC_n\:|n+1;(0,0)\rangle\langle n;(0,0)|\:e^{i\frac{\theta_0}{\hbar}J^2}+
	\right.\nonumber\\
	&&\left.{h'}^*(t)\:e^{-i\frac{\theta_0}{\hbar}J^2}   \sum_nC_n\:
	|n-1;(0,0)\rangle\langle n;(0,0)|\:e^{i\frac{\theta_0}{\hbar}J^2}\right]\;,\nonumber\\
	&=&\frac{1}{2}\left[ {h'}(t)\:e^{-i\frac{\theta_0}{\hbar}J^2}J^+
	e^{i\frac{\theta_0}{\hbar}J^2}+{h'}^*(t)\:e^{-i\frac{\theta_0}{\hbar}J^2} 
	J^-e^{i\frac{\theta_0}{\hbar}J^2}\right] \;,\nonumber\\
	&=& {\rm Re}[{h'}(t)]\: (\cos\theta_0\:J^1-\sin\theta_0\:J^3)-
	{\rm Im}[{h'}(t)]\:J^2\;,
	\label{h^1-2}
	\end{eqnarray}
where
	\begin{equation}
	{h'}(t):=\: e^{i[bt+\gamma-\varphi]}
	(i\:\dot\theta+\sin\theta\:\dot\varphi)\;,
	\label{h'}
	\end{equation}
$\theta_0:=\theta(0)$, and ``Re'' and ``Im'' stand for the real part and ($-i$ times) 
the imaginary part of the argument. Furthermore, in the derivation of 
(\ref{h^1-2}) use is made of  Eqs.~(\ref{h^1-spin}), (\ref{3.2}), (\ref{j+-}), 
$\varphi(0)=0$, and the identities:
	\begin{equation}
	e^{-i\frac{\theta_0}{\hbar}J^2}J^{\pm}e^{i\frac{\theta_0}{\hbar}J^2}=
	e^{-i\frac{\theta_0}{\hbar}J^2}J^1e^{i\frac{\theta_0}{\hbar}J^2}\pm iJ^2
	=\cos\theta_0\:J^1-\sin\theta_0\:J^3\pm iJ^2\;.
	\nonumber
	\end{equation}

Next consider the case in  which the magnetic field performs
a simultaneous precession and nutation, i.e.,
	\begin{equation}
	C:~\varphi=\omega t\;,~~~\cos\theta=\cos\theta_0+\eta\sin(l\omega t)\;.
	\label{pn}
	\end{equation}
Here $\omega$ is the frequency of precession, 
and $|l|\omega$ and $\eta$ are the frequency and amplitude of the nutation,
respectively. Clearly $0\leq \eta<1\pm\cos\theta_0$ and $l$ is an
integer. 

The computation of $\gamma$ is then straightforward:
	\begin{equation}
	\gamma(t)=(1-\cos\theta_0)\omega t-\frac{\eta}{l}[1-\cos(l\omega t)]\;.
	\label{gamma-pn}
	\end{equation}
Hence for $\tau=T:=2\pi/\omega$, one recovers Berry's result:
	\begin{equation}
	\gamma_n(T)=-2\pi n(1-\cos\theta_0)=-n\Omega(C)\;,
	\label{gamma-n-pn}
	\end{equation}
where $\Omega(C)$ is the solid angle subtended by the curve $C$. 
A simple consequence of (\ref{gamma-n-pn}) is that the adiabatic
approximation does not detect the effects of the nutation. In fact, one
has:
	\begin{equation}
	U^{(0)}(T)=\sum_n e^{-i[bT+2\pi (1-\cos\theta_0)]n}
	|n;0\rangle\langle n;0|\:=\:e^{-i\frac{\theta_0}{\hbar}J^2}
	e^{-\frac{i}{\hbar}[bT+2\pi (1-\cos\theta_0)]J^3}
	e^{i\frac{\theta_0}{\hbar}J^2}\;.
	\label{u0-pn}
	\end{equation}

The first non-adiabatic correction  $U^{(1)}(T)$ to (\ref{u0-pn}) is obtained
by  integrating $H'(t)$ of Eq.~(\ref{h^1-2}). This involves the evaluation of
	\begin{equation}
	I(T):=\int_0^T h'(t)dt=\int_0^T dt\: e^{i[bt+\gamma-\varphi]}
	(i\:\dot\theta+\sin\theta\:\dot\varphi)\;.
	\label{i}
	\end{equation}
In order to give a closed expression for this integral and subsequently 
$U^{(1)}(T)$, one can expand the integrand in powers of $\eta$. This leads
to a lengthy calculation which results in:
	\begin{eqnarray}
	{\rm Re}[I(T)]&=&\zeta\:\left\{
	\sin\theta_0\sin(2\pi/\zeta)+\left[ (\frac{1-\cos(2\pi/\zeta)}{1-(l\zeta)^2})
	(\frac{1}{\sin\theta_0}+\cot\theta_0\:\zeta-\sin\theta_0\:\zeta^2)\right]
	l\eta+\right.\nonumber\\
	&&\left.\left[ \frac{2\cos\theta_0\sin(2\pi/\zeta)\zeta}{\sin^3\theta_0
	[1-4(l\zeta)^2]}	 \right](l\eta)^2\right\}+{\cal O}(l\eta^2)
	\;,
	\label{re-i}\\
	{\rm Im}[I(T)]&=&\zeta\:\left\{
	\sin\theta_0[1-\cos(2\pi/\zeta)]+\left[(-\frac{1}{\sin\theta_0}-
	\cot\theta_0\zeta+\sin\theta_0\zeta^2)(\frac{\sin(2\pi/\zeta}{1-(l\zeta)^2})
	\right]\: l\eta+\right.\nonumber\\
	&&\left.\left[ \frac{2\cos\theta_0[1-\cos(2\pi/\zeta)]\zeta}{\sin^3\theta_0
	[1-4(l\zeta)^2]}\right]\:(l\eta)^2\right\}+{\cal O}(l\eta^2)
	\;,
	\label{im-i}
	\end{eqnarray}
where $\zeta:=\omega/(b-\omega\cos\theta_0)$. In terms of ${\rm Re}[I(T)]$
and ${\rm Im}[I(T)]$, $U^{(1)}(T)$ is expressed in the form:
	\begin{eqnarray}
	U^{(1)}(T)&=&-\frac{i}{\hbar} \:U^{(0)}(T)\:\left\{
	 {\rm Re}[{I}(T)]\: (\cos\theta_0\:J^1-\sin\theta_0\:J^3)-
	{\rm Im}[{I}(T)]\:J^2\right\}\;,\nonumber\\
	&=&-\frac{i}{\hbar} \;e^{-i\frac{\theta_0}{\hbar}J^2}
	e^{-\frac{i}{\hbar}[bT+2\pi (1-\cos\theta_0)]J^3}
	\left\{ {\rm Re}[I(T)]\:J^1-{\rm Im}[I(T)]\:J^2\right\}
	e^{i\frac{\theta_0}{\hbar}J^2}\:.
	\label{u^1-pn}
	\end{eqnarray}

In particular, consider the special case, $\theta_0=\pi/2$. Then
$\zeta=\omega/b$ and Eqs.~(\ref{u0-pn}) and (\ref{re-i}-\ref{u^1-pn}) 
reduce to:
	\begin{eqnarray}
	U^{(0)}(T)&=&e^{-i\frac{\pi}{2\hbar}J^2}
	e^{-\frac{i}{\hbar}(bT+2\pi )J^3}
	e^{i\frac{\pi}{2\hbar}J^2}\:=\:e^{-i2\pi  j}e^{-i\frac{\pi}{2\hbar}J^2}
	e^{-\frac{i}{\hbar}bTJ^3}e^{i\frac{\pi}{2\hbar}J^2}
	\;,
	\label{u0-pn-0}\\
	{\rm Re}[I(T)]&=&\zeta\:\left\{
	\sin(2\pi/\zeta)+\left[ (\frac{1-\zeta^2}{1-(l\zeta)^2})
	[1-\cos(2\pi/\zeta)]	\right]l\eta\right\}+{\cal O}(l\eta^2)
	\;,
	\label{re-i-0}\\
	{\rm Im}[I(T)]&=&\zeta\:\left\{
	1-\cos(2\pi/\zeta)+\left[(-\frac{1-\zeta^2}{1-(l\zeta)^2})\sin(2\pi/\zeta)
	\right]l\eta\right\}+{\cal O}(l\eta^2)
	\;,
	\label{im-i-0}\\
	U^{(1)}(T)&=&-\frac{i}{\hbar} \:e^{-i2\pi  j}
	e^{-i\frac{\pi}{2\hbar}J^2}
	e^{-\frac{i}{\hbar}bTJ^3}
	\left\{ {\rm Re}[I(T)]\:J^1-{\rm Im}[I(T)]\:J^2\right\}
	e^{i\frac{\pi}{2\hbar}J^2}\:.
	\label{u^1-pn-0}
	\end{eqnarray}
The adiabaticity parameter $\nu$ of Eq.~(\ref{slow}) can also be
easily calculated in this case:
	\begin{equation}
	\nu=c\sqrt{1+(l\eta)^2}\;\zeta+{\cal O}(l\eta^2)\;,
	\label{slow-0}
	\end{equation}
where $c:=C_{-j}/2=\sqrt{j(j+1/2)}$. The appearance of $\zeta$
in this expression, is an indication of the fact that the leading
order term in $U^{(1)}(T)$ is of order $\nu$. In fact, for the two
extreme cases: $l\eta\ll 1$ and $l\eta\gg 1$ (with $l\eta^2\ll 1$
so that one can neglect ${\cal O}(l\eta^2)$) this is manifestly
seen.

As indicated by Eqs.~(\ref{u0-pn-0}-\ref{u^1-pn-0}), the nutation of
the direction of the magnetic field only contributes to the
non-adiabatic part of the evolution operator.

\section{Summary and Conclusion}
In this article, I have tried to address some of the basic issues regarding
the meaning of the adiabatic approximation and the implications
of their resolution  to the phenomenon of the geometric
phases. A precise definition of an adiabaticity parameter, although being 
rather unimportant in practice, provides an objective criterion
for the applicability of the adiabatic approximation. It also allows one 
to deal with some of the misconceptions concerning the subject. 
Specifically, it is shown that there is no relation between the length of the
duration of the evolution and the validity of the adiabatic approximation. 
Furthermore for periodic Hamiltonians, the relation between the
adiabaticity of the evolution and the period of the Hamiltonian is
shown to be not a simple proportionality. For example for the system
described by Eqs.~(\ref{3.1}) and (\ref{pn}), both the period $T$ and the
nutation parameter $l\eta$ might be arbitrarily large, in which case the
adiabatic approximation is not valid. An example of a periodic system
with an arbitrary small period for which the adiabatic approximation
is exact, is a system with constant energy eigenstates but time
dependent energy eigenvalues. A typical example of such a
system is a magnetic dipole in a magnetic field whose magnitude 
is rapidly changing but its direction is fixed. Clearly, in this case,
the adiabaticity parameter $\nu$ vanishes identically and the
adiabatic approximation is valid exactly.

A perturbative expansion of the evolution operator $U(\tau)$
is also proposed. This expansion generalizes the adiabatic
approximation and computes the non-adiabatic corrections
to the adiabatically approximate expression for $U(\tau)$.
Furthermore, it separates the adiabatic and non-adiabatic parts
of $U(\tau)$ and yields the non-adiabatic part as a time-ordered
exponential defined by a Hermitian operator $H'(t)$. In a basis 
which diagonalizes the initial Hamiltonian, $H'(t)$ is off-diagonal. 
Moreover, the dominant contribution to this operator and 
consequently to the non-adiabatic part of the evolution
operator comes from the nearest energy levels. Particularly
interesting is to view $H'(t)$ as a transformed Hamiltonian. This
corresponds to a time-dependent unitary transformation of the
state vectors which undoes the adiabatic part of the evolution
operator. 

The subject of the general time-dependent unitary transformations
of the state vectors has also been reviewed and a simple application
of such a transformation is used to yield a splitting of the total phase
of a cyclic state into its dynamical and geometric parts for a periodically
changing Hamiltonian. This is followed by a discussion of physically
equivalent quantum systems and the conclusion that unlike the 
dynamical  (total) phase of a cyclic state,  the difference of the
dynamical (total) phases of two cyclic states is a physical quantity. 

Finally, the theoretical developments have been applied to the analysis
of a quantum system consisting of a magnetic dipole subject to a 
magnetic field. In particular, an explicit formula for the operator $H'(t)$
has been derived for the general case and the case of a precessing and
nutating magnetic field has been studied in detail.

\section*{Acknowledgements}
I would like to thank Bahman Darian and Rouzbeh Allahverdi for
interesting discussions, and acknowledge the support of the
Killam Foundation of Canada.

\end{document}